\documentclass[twocolumn,aps,pra,showpacs,floatfix,superscriptaddress]{revtex4-1}
\usepackage{graphicx} 
\usepackage{amsmath}
\usepackage{bm}
\usepackage{color}
\usepackage{dcolumn}

\begin{document}

\newcommand{\NZIAS}{
Centre for Theoretical Chemistry and Physics,
The New Zealand Institute for Advanced Study,
Massey University Auckland, Private Bag 102904, 0745 Auckland, New Zealand}

\newcommand{\UNSW}{
School of Physics, University of New South Wales, Sydney 2052, Australia}

\title{Sensitivity to $\alpha$-variation in ultracold atomic-scattering experiments}

\author{A. Borschevsky}
\affiliation{\NZIAS}
\affiliation{\UNSW}

\author{K. Beloy}
\affiliation{\NZIAS}
\affiliation{\UNSW}

\author{V. V. Flambaum}
\affiliation{\UNSW}
\affiliation{\NZIAS}

\author{P. Schwerdtfeger}
\affiliation{\NZIAS}

\date{\today}

\begin{abstract}
We present numerical calculations for cesium and mercury to estimate the sensitivity of the scattering length to the variation of the fine structure constant $\alpha$.
The method used follows ideas Chin and Flambaum [Phys. Rev. Lett.~{\bf 96}, 230801 (2006)], where the sensitivity to the variation of the electron to proton mass ratio, $\beta$, was
considered. We demonstrate that for heavy systems, the sensitivity to variation of $\alpha$ is of the same order of magnitude as to variation of $\beta$. 
Near narrow Feshbach resonances the enhancement of the sensitivity may exceed nine orders of magnitude.
\end{abstract}


\pacs{34.50.Cx, 34.20.Cf, 06.20.Jr}
\maketitle

\newcommand{\cmt}[1]{[\![#1]\!]}

\section{Introduction}

Theories unifying gravity with other interactions suggest the possibility of
spatial and temporal variation of fundamental physical constants, such as the
fine structure constant, $\alpha=e^{2}/\hbar c$, and the electron to proton
mass ratio, $\beta=m_{e}/m_{p}$ \cite{Uza03}. The search for such variations has
received considerable interest in recent years, and is being conducted using a
wide variety of methods (see, e.g., reviews \cite{Fla07,Fla08}). Two major directions of search for
variation in $\alpha$ and in $\beta$ 
are observational studies, such as
analysis of high resolution spectroscopy from dark clouds in interstellar space
(see, e.g., Ref.~\cite{MurWebFla07,WebKinMur10,DzuFlaWeb99,DzuFlaWeb99_2,LevMolLap10,
LevLapHen10} and references
therein), and laboratory research, including frequency
comparison of optical and microwave atomic clocks over extended periods of
time (see, e.g., \cite{PreTjoMal95,MarPerSan03,ForAshBer07,Ros08,DzuFlaWeb99,DzuFlaWeb99_2}). 

Great advantage may be gained in these searches by focusing on systems that
have enhanced sensitivity to variation of $\alpha$ or $\beta$. For spectroscopy experiments, such
an enhancement may occur due to a quasi-degeneracy of two close lying levels
of different nature, found in some atoms
\cite{DzuFlaWeb99,DzuFlaWeb99_2,DzuFlaMar03,DzuFla05,NguBudLam04}, molecules \cite{FlaKoz07,BelBorSch10}, 
and nuclei \cite{PeiTam03,Fla06,LitFelDob09}. Chin and Flambaum pointed out that a large enhancement 
(on the order of $10^{9}-10^{12}$) could also be achieved in ultracold atomic collision experiments near Feshbach resonances \cite{ChiFla06}.
In their work, the sensitivity of the scattering length to variations in the electron to proton mass ratio was investigated.
Here, we extend this work by further considering
sensitivity to the fine structure constant.

The $s$-wave scattering length associated with the collision of two neutral atoms may be written as \cite{GriFla93}
\begin{equation}
a=\bar{a}\left[1-\mathrm{tan}\left(\phi-\frac{\pi}{8}\right)\right],
\label{Eq:scattlength}
\end{equation}
where $\bar{a}$ is the mean scattering length and $\phi$ is the scattering phase shift. The mean scattering 
length is further given by $\bar{a}=c(2\mu C_6/\hbar^2)^{1/4}$, with $c$ being a constant ($\approx0.47799$), 
$\mu$ the reduced mass, and $C_6$ the van der Waals coefficient characterizing the long-range interatomic interaction. 
Employing the semi-classical WKB approximation, the phase shift reads \cite{GriFla93}
\begin{equation}
\phi=\int_{r_i}^{\infty}\hbar^{-1}\sqrt{-2\mu [V(r)-V(\infty)]}dr,
\label{Eq:phi}
\end{equation}
where $V(r)$ is the molecular potential at internuclear separation $r$
and the limit $r_i$ corresponds to the inner turning point for zero kinetic energy of the colliding atoms, i.e. $V(r_i)=V(\infty)$
at the onset of the repulsive wall.

From Eq.~(\ref{Eq:scattlength}), we see that the scattering length diverges at values of $\phi=(n+\frac{5}{8})\pi$, with $n$ being an integer. 
Such divergences may lead to enhanced sensitivity of the scattering length $a$ to variation of the fundamental constants: 
a small change in the phase shift $\phi$ could produce a significant change in the scattering length. However, the main enhancement is due to
the large ratio $(\mu/m_e)^{1/2}\sim 10^2-10^3$,
where $\mu$ is the reduced mass and $m_e$ is the  electron mass.
This large parameter justifies the applicability of WKB approximation,
Eq.~(\ref{Eq:phi}), to the motion of a heavy atom in the inter-atomic potential,
which has an electron-volt scale,  and also  makes the phase very large.
As a result, a tiny {\it fractional} change in the phase 
$\phi$ produces a dramatic change in 
 $\mathrm{tan}\left(\phi-\frac{\pi}{8}\right)$ and the scattering length in
 Eq.~(\ref{Eq:phi}). Therefore, the effects of the variation of the fundamental
constants manifest themselves mainly via a change in the phase $\phi$.

From the explicit dependence of $\phi$, Eq.~(\ref{Eq:phi}), 
on the reduced mass $\mu$, it follows that a variation in $\beta$ produces an associated change in $\phi$ according to the simple relation \cite{ChiFla06}
\begin{equation}
\frac{\delta\phi}{\phi}=-\frac{1}{2}\frac{\delta\beta}{\beta}.
\label{Eq:dphibeta}
\end{equation}

Here we consider the change of the scattering phase shift $\phi$ with respect to variation of the
fine structure constant $\alpha$. The $\alpha$-dependence arises from within the molecular potential $V(r)$
due to relativistic effects, and the extraction of this dependence therefore requires \emph{ab-initio} 
relativistic computations of the molecular potential. 
We compute the potential energy curve for the $^1\Sigma^+_g$ ground state of the Cs$_2$ dimer, and use the previously calculated 
relativistic and non-relativistic $^1\Sigma^+_g$ ground state potential energy curves for Hg$_2$ 
published by Pahl \textit{et al}.~\cite{PahFigBor11,PahFigThi10},
obtained at the coupled cluster level in the complete basis set limit.
With these potential energy curves we are able to obtain reasonable estimates for the sensitivity 
of $\phi$ to $\alpha$-variation for Cs$_2$ and Hg$_2$. We present our results in terms of the factor $K_\alpha$, where
\begin{equation}
\frac{\delta\phi}{\phi}=K_\alpha\frac{\delta\alpha}{\alpha}+K_\beta\frac{\delta\beta}{\beta},
\label{Eq:dphialphabeta}
\end{equation}
and $K_\beta=-1/2$. From this one can derive the change in the scattering length (Eq.~(\ref{Eq:scattlength})) 
with respect to variation of $\alpha$ and $\beta$ \cite{ChiFla06},
\begin{equation}
\frac{\delta a}{a}=\frac{N\pi}{2}\frac{(a-\bar{a})^2+\bar{a}^2}{a\bar{a}}\left(K_\alpha\frac{\delta \alpha}{\alpha}
+K_\beta\frac{\delta \beta}{\beta}\right).
\end{equation}
The above relation relies on two assumptions. First the potential should be deep enough or the reduced mass
large enough (as for Cs$_2$ or Hg$_2$) to support a large number of bound vibrational states, $N$. 
Then the phase can be estimated as $\phi \approx N\pi \gg \pi$ \cite{GriFla93}. 
The second assumption is in neglecting the variation of the mean scattering length, $\bar{a}$, which is justified 
by the fact that it does not benefit from the enhanced sensitivity of the tangent function appearing in Eq.~(\ref{Eq:scattlength}).
We then go on to extend our findings to the case of Feshbach resonances, where even greater enhancement of sensitivity
to variation of fundamental constants can be expected \cite{ChiFla06}. 
The dynamic formation of quasi-bound Cs$_2$ molecules near narrow Feshbach resonances has been observed by Chin \textit{et al}.~\cite{ChiCheKer03}.

The scattering length has dimension of length; hence we have to specify the units we use. It is convenient to use atomic units (of energy, length, etc.). 
In these units the dependence on $\alpha$ appears due to the relativistic corrections (i.e., terms containing $1/c$). 
Note that use of different units would not change our conclusions since the variation of the ratio of commonly used units 
(e.g., the Bohr radius and meter) is not enhanced, while the variation of the ratio of the scattering length to the Bohr radius 
is enhanced many orders of magnitude; for a detailed explanation see Ref.~\cite{ChiFla06}. 


\section{Method and results}



\begin{figure}[t]
\begin{center}
\includegraphics*[scale=0.75]{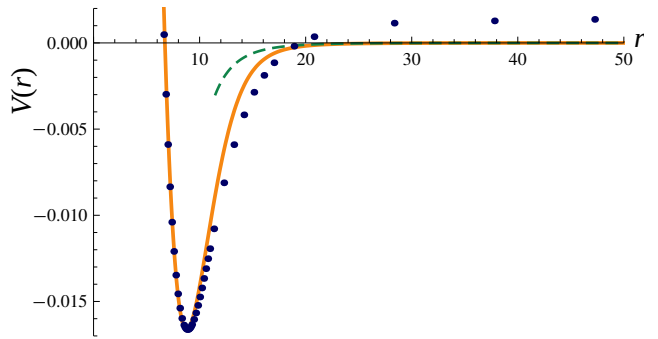}
\end{center}
\caption{(color online) 
Potential energy curve for the $^1\Sigma^+_g$ ground state of Cs$_2$.
The circles correspond to values calculated as described in the text and the solid line corresponds to the $V_\mathrm{MLR3}(r)$ potential of Ref.~\cite{CoxHaj10}. 
The dashed line represents the asymptotic part of the true potential energy curve, $-C_6/r^6$.
All units are atomic units.}
\label{Fig:Cs2curve}
\end{figure}

The first system of interest here is the cesium dimer. A fully analytical potential energy curve for the Cs$_2$ ground state 
has recently been given by Coxon and Hajigeorgiou \cite{CoxHaj10}. This curve employs several fitting parameters to accurately 
reproduce experimental spectroscopic data over 99.24\% of the well depth. Additionally, this curve has the 
appropriate asymptotic behavior describing the neutral atoms interacting at long-range, 
namely $V(r)\rightarrow V(\infty)-C_6/r^6$ as $r\rightarrow\infty$.
This analytical potential energy curve, denoted as MLR3 by the authors, is displayed in Fig.~\ref{Fig:Cs2curve}. 
Performing the integration in Eq.~(\ref{Eq:phi}), we find the scattering phase shift
\begin{equation}
\phi_\mathrm{MLR3}=487.
\label{Eq:phiMLR3}
\end{equation}

We have computed the molecular potential curve 
at incremental values of $r$ using the relativistic molecular 
four-component program package DIRAC~\cite{DIRAC10}. 
Details of this calculation are saved for the Appendix.
Our resulting values are overlaid on the MLR3 potential curve in Fig.~\ref{Fig:Cs2curve}.
From inspection of this figure, we note a good agreement between our computed values 
and $V_\mathrm{MLR3}(r)$ in the vicinity of the equilibrium distance $r_e$ as well as at shorter distances. 
From the equilibrium distance we continue to have reasonable agreement up to $r\approx1.5r_e\approx14$~a.u., 
beyond which we compute values quite different to the asymptotic behavior of $V_\mathrm{MLR3}(r)$.
Moreover, the well depth is 10\% too large. This is due to limitations in the basis set applied and 
the neglect of higher order contributions in our coupled-cluster expansion. Further, the Breit interaction may
also change the potential energy curve in this region.

In principle, we may calculate $\phi$ using the potential energy curve determined from a fit of our computed data 
points, $V_\mathrm{CCSD}(r)$. We find, however, that this gives an entirely unsatisfactory result due to the inaccurate 
long-range behavior of this curve. Firstly, the excessive well-depth implies that integration over $r$ in even the 
short-range is erroneous (the integrand of Eq.~(\ref{Eq:phi}) contains the difference $V(r)-V(\infty)$). 
Secondly, the wrong asymptotic behavior of $V_\mathrm{CCSD}(r)$ implies that a considerable amount 
of phase is inappropriately accumulated in the long-range beyond $r\approx14$ a.u. (the slow convergence is 
effectively magnified by the square root in the integrand of Eq.~(\ref{Eq:phi})).
Clearly, to obtain a reasonable value of $\phi$ it is necessary to satisfactorily treat the long-range 
part of the potential. We choose to do so by replacing the erroneous long-range part of our computed curve with the proper asymptotic behavior, $-C_6/r^6$,
at a suitably chosen switching point $r_a$. There is a certain degree of arbitrariness in this choice; 
we set $r_a=14$, as at this point the deviation of the MLR3 curve from the
asymptotic curve, $-C_6/r^6$, is roughly equal to its deviation from the computed energy curve, but of an opposite sign, leading 
to error cancellation. Taking the experimental 
potential depth and $C_6$ parameters \cite{CoxHaj10}, we compute the scattering phase shift to be
\begin{equation}
\phi=477,
\label{Eq:phiref}
\end{equation}
this result being 2\% lower than the value obtained from the accurate MLR3 curve, Eq.~(\ref{Eq:phiMLR3}).

We recompute the potential energy curve 
for different values of $\alpha$ in the neighborhood of $\alpha_0=1/137.036$, and obtain the scattering phase 
using the method outlined above. It should be noted that the tail part of the potential has a considerable
contribution to the calculated phase (for the MRL3 curve, as much as 28\% of the phase accumulates 
after $r=14$ a.u.). Thus it is important to also take into account the variation of the
$C_6$ parameter with respect to variations in $\alpha$. We obtain its dependence on $\alpha$ in the Appendix, and
adjust the integration parameters accordingly.
The potential well depth,
$D_e$, is rescaled using the following relation: 
$
D_e(\alpha)=[D_e^\mathrm{expt.}/D_e^\mathrm{CCSD}(\alpha_0)] D_e^\mathrm{CCSD}(\alpha),
$
where $D_e^\mathrm{CCSD}(\alpha)$ is the calculated depth of the potential energy well for a given value of $\alpha$.
By taking the numerical derivative of $\phi$ with respect to $\alpha$ we are 
then able to obtain our estimated factor $K_\alpha$ in Eq.~(\ref{Eq:dphialphabeta}).

A different strategy was adopted for the mercury dimer. High accuracy calculations of both relativistic and non-relativistic
potential energy curves were recently carried out with the correct long-range behavior from
symmetry-adapted perturbation theory \cite{PahFigBor11,PahFigThi10}. We integrate over these potentials to 
obtain $\phi_{\alpha=\alpha_0}$ and $\phi_{\alpha=0}$ for relativistic and non-relativistic energy curves, respectively. 
Assuming linear dependence of molecular properties on $\alpha^2$, we use the two points to derive $K_\alpha$ for Hg$_2$. 
Note that we verified that the assumption of the linear dependence on $\alpha^2$ works very well for Cs$_2$. 
In mercury the non-linear corrections will be larger; however, they should not significantly exceed the 
natural relativistic parameter $Z^2 \alpha^2$, $Z$ being the atomic number.

The resulting values for the two systems are

\begin{equation*}
K_\alpha(\mathrm{Cs})
=0.04,
\qquad\quad
K_\alpha(\mathrm{Hg})
=-0.3.
\end{equation*}

A typical value of relativistic corrections is $Z^2 \alpha^2$, which is equal to 0.34 for Hg. This naturally 
explains the magnitude of $K_\alpha$ for this system. For Cs the
result is smaller, due to an accidental cancellation between the contributions of the main and the asymptotic parts 
of the potential energy curve. It should be stressed here that the number 
of approximations employed in this work do not allow us to set reliable error limits on the obtained values of 
$K_\alpha$. However, performing test calculations with reasonable variation of the free parameters (e.g., $r_a$ and $C_6$) does not change the final result by more than 
30\%, which gives us a rough error estimate. We can nevertheless conclude that for heavy 
relativistic systems the sensitivity of the scattering phase to $\alpha$ is of the same order of magnitude as its 
sensitivity to the electron to proton mass ratio ($K_\beta=-0.5$). This means that we have strongly enhanced 
sensitivity of the scattering length to the variation of $\alpha$
and may use the equations presented in the Introduction to estimate this sensitivity. 

A general estimate of the scattering length sensitivity for arbitrary atoms is
\begin{equation*}
\frac{\delta a}{a} \sim \pi N \left(K_\beta \frac{\delta \beta}{\beta}+K_\alpha\frac{\delta \alpha}{\alpha}\right),
\end{equation*}
where
\begin{equation*}
K_\beta=-0.5,\,\,\, K_\alpha \sim Z^2 \alpha^2.
\end{equation*}
The enhancement is due to the large number $N$ of the vibrational levels in the interatomic potential.
For Cs, where $N=150$, $a=280$, and ${\bar a}=95$ \cite{ChiFla06},the enhancement coefficient is given by 
\begin{equation*}
\frac{\delta a}{a}=800(0.5 \frac{\delta \beta}{\beta}- 0.04\frac{\delta \alpha}{\alpha}).
\end{equation*}

More accurate result requires knowledge of the experimental value of the scattering length. Unfortunately, the scattering length
for Hg$_2$ is not available from experiment.

\section{Extension to Feshbach Resonances}
In real scattering systems, the molecular potential has multiple channels. When a bound
state in a closed channel near the dissociation limit is tuned close to the scattering state in an open channel, 
Feshbach coupling between the two channels can lead to a resonant enhancement of the scattering length (following Ref.~\cite{ChiFla06}):
\begin{equation}
A=a
\left(
1+\frac{\Delta E}{E_o-E_m}
\right),
\label{Eq:A}
\end{equation}
where $A$ is the scattering length near the resonance, $a$ is the off-resonant scattering 
length, $\Delta E$ characterizes the Feshbach coupling strength,
 $E_o$ is the energy of the system (in the case of cold scattering we can put
 $E_o=0$), and $E_m$ is the energy of the bound state. The main effect comes
 from the variation of the small energy denominator.
Variation of Eq.~(\ref{Eq:A}) is then \cite{ChiFla06}
\begin{equation*}
\frac{\delta A}{A}=
\frac{\delta a}{a}
+\frac{A-a}{A}
\frac{\delta E_m}{E_o-E_m},
\end{equation*}
where the small effect of the  variation of $\Delta E$ is neglected.

To explore the dependence of $A$ on $\alpha$ and $\beta$, we express $\delta A/A$ in a manner similar to Eq.~(\ref{Eq:dphialphabeta}),
\begin{equation}
\frac{\delta A}{A}=R_\alpha\frac{\delta \alpha}{\alpha}
+R_\beta\frac{\delta \beta}{\beta}.
\end{equation}

Chin and Flambaum~\cite{ChiFla06} have derived an expression of the sensitivity of $A$ with respect to the 
electron to proton mass ratio by exploring the dependence of the bound state energy $E_m$ on $\beta$ 
and by using the following relation between the phase, $\phi_M$, and $E_m$, derived from the WKB approximation similar to Eq.~(\ref{Eq:phi}),
\begin{equation}
\phi_M=\int_{r_i}^{r_o}\hbar^{-1}\sqrt{2\mu\left[E_m-V_c(r)\right]}dr=M\pi.
\label{Eq:phi2}
\end{equation}
Here, $M$ is the vibrational quantum number of the bound state, and $r_i$ and $r_o$ are the classical 
inner and outer turning points, respectively, for a total energy $E_m$.
They further obtained \cite{ChiFla06},
\begin{equation*}
R_\beta=
\frac{M}{2}
\frac{(A-a)^2}{Aa}
\frac{1}{\rho(E_m)\Delta E},
\end{equation*}
where $\rho(E_m)$ is the density of states at the energy $E_m$ and can be estimated from the 
energy splitting, $D$, between adjacent vibrational levels, $\rho(E_m)\approx 1/D$.
From similarity of expressions (\ref{Eq:phi2}) and (\ref{Eq:phi}),
we may provide a rough estimate, $R_\alpha/R_\beta\sim K_\alpha/K_\beta \sim Z^2\alpha^2$.
This means that for the heavy atoms the enhancement coefficients
for the effects of $\mu$ and $\alpha$ variations are comparable.

According to Ref. \cite{ChiFla06}, this enhancement near narrow Feshbach
resonances (e.g., near $g$-wave resonance in Cs$_2$) may reach $10^9 -10^{12}$.
Recently, a method for measurement of the scattering length to the accuracy
$10^{-6}$ was proposed \cite{HartXuLeg07}. A combination of these two ideas makes the
repeating measurements  of the scattering length a promising method to study
the variation of the fundamental constants in the laboratory.
Here we mention that the creation of ultracold $^{133}$Cs$_2$ from Feshbach resonances from
Bose-Einstein condensates was demonstrated already by Herbig \textit{et al.}~\cite{HerKraWeb03}.
In principle, cosmic data about scattering lengths and chemical reaction rates in the early
Universe may be used too.

\section{Appendix: Computational Details}

The potential energy curves of Cs$_2$ were calculated using the 2010 version of the DIRAC program
\cite{DIRAC10}. In order to reduce the computational effort, the
four-component Dirac-Coulomb Hamiltonian was replaced by the infinite order
two-component relativistic Hamiltonian obtained after the Barysz-Sadlej-Snijders
(BSS) transformation of the Dirac Hamiltonian in a finite basis set
\cite{IliJenKel05}. This approximation includes both scalar and spin-orbit
relativistic effects to infinite order, and is one of the most computer time efficient
and accurate approximations to the four-component Dirac-Coulomb Hamiltonian.

For Cs$_{2}$ electron correlation was
taken into account using closed-shell single-reference coupled-cluster
theory including single and double excitations (CCSD). The Faegri dual basis set
\cite{Fae01} was used, consisting of 23\textit{s}%
19\textit{p}13\textit{d}4\textit{f}2\textit{g} Gaussian orbitals. 
Virtual orbitals with energies above 45 a.u.~were omitted, and the 54 outer-most electrons 
in the valence space were correlated.
The molecular energy curves were recomputed for different values of $\alpha$, 
and, together with the computed sensitivity of $D_e$ and $C_6$ to $\alpha$, are used to obtain the dependence of the
scattering phase on the fine structure constant.
 
In order to obtain the dependence of the $C_6$ coefficients on $\alpha$ we use the following scheme.
The van der Waals $C_6$ coefficient may be written in atomic units as \cite{CasPol48,Lon30,Pow01}
\begin{equation}
C_6=\frac{3}{\pi}\int_0^\infty d\omega [\alpha_P(i\omega)]^2,
\label{Eq:C6}
\end{equation}
where $\alpha_P(\omega)$ is the dynamic scalar polarizability. For atomic state $n$, the polarizability is given by \cite{DerJohSaf99}
\begin{equation*}
\alpha_P(\omega)=\frac{2}{3}\sum_{n^\prime}\left|\langle n^\prime|\mathbf{D}|n\rangle\right|^2\left(\frac{\Delta_{n^\prime n}}{\Delta^2_{n^\prime n}-\omega^2}\right),
\end{equation*}
with $\mathbf{D}$ the electric dipole operator and $\Delta_{n^\prime n}\equiv E_{n^\prime}-E_n$ being the energy difference between states. If  a single intermediate state $n^\prime$ dominates the summation, we may then estimate $C_6$ by
\begin{eqnarray}
C_6
&\approx&\frac{3}{\pi}\left(\frac{2}{3}\right)^2\frac{\left|\langle n^\prime|\mathbf{D}|n\rangle\right|^4}{\Delta_{n^\prime n}^2}\int_0^\infty d\omega\left(\frac{\Delta_{n^\prime n}^2}{\Delta^2_{n^\prime n}+\omega^2}\right)^2
\nonumber\\
&\approx&\frac{3}{4}\left|\alpha_P(0)\right|^2\Delta_{n^\prime n},
\label{Eq:C6approx}
\end{eqnarray}
where we use $\int_0^\infty(1+x^2)^{-2}dx=\pi/4$.

For the Cs ground state, $n=6s_{1/2}$, the majority (96\%) of the of the total polarizability is due to the $n^\prime=6p_{1/2,3/2}$ intermediate states \cite{DerJohSaf99}. 
The fractional change in the $C_6$ coefficient can then be estimated using
\begin{equation}
\frac{\delta C_6}{C_6}
\approx2\frac{\delta\alpha_P(0)}{\alpha_P(0)}+\frac{\delta \Delta}{\Delta},
\label{Eq:C6Frac}
\end{equation}
where the first term is the fractional variation in the static polarizability, and the second term represents the fractional variation in the first allowed transition energy. 
It should be noted that for Cs, 
we take into account both the 
$6p_{1/2,3/2}$ states, using the weighted average
$\Delta=(\Delta_{1/2}+2\Delta_{3/2})/3$ of the excitation energies.

The static polarizability and the low excited energy states were calculated using the Fock space coupled cluster approach (FSCC). Due to the high sensitivity 
of these properties to diffuse basis functions, a slightly larger basis was used, consisting of  25\textit{s}%
21\textit{p}14\textit{d}6\textit{f}3\textit{g} orbitals. The polarizability was obtained using the finite field approach, similar to
Ref.~\cite{PerBorEli08}. The calculations were performed for different values of $\alpha$, in the vicinity of $\alpha_0$, and their dependence on the fine 
structure constant was determined using numerical differentiation. Our calculated static polarizability for Cs with 396.5 a.u.~is in good agreement
with the most precise values from two other groups (399.9 a.u.~\cite{DerJohSaf99} and 
396.0 a.u.~\cite{LimSchMet05}) and with experimental measurements (401.0$\pm$0.6 a.u.~\cite{AmiJasGou03}).
Using Eq.~(\ref{Eq:C6Frac}), we obtain 
\begin{equation*}
\frac{\delta C_6}{C_6}(\mathrm{Cs})\approx-0.572\frac{\delta\alpha}{\alpha}.
\end{equation*}

\section{Acknowledgements}
This work was supported by the Marsden Fund, administered by the Royal Society of New Zealand, and the Australian Research Council.
We thank E. Pahl (Auckland) for useful discussions and help.


\begin{thebibliography}{39}%
\makeatletter
\providecommand \@ifxundefined [1]{%
 \@ifx{#1\undefined}
}%
\providecommand \@ifnum [1]{%
 \ifnum #1\expandafter \@firstoftwo
 \else \expandafter \@secondoftwo
 \fi
}%
\providecommand \@ifx [1]{%
 \ifx #1\expandafter \@firstoftwo
 \else \expandafter \@secondoftwo
 \fi
}%
\providecommand \natexlab [1]{#1}%
\providecommand \enquote  [1]{``#1''}%
\providecommand \bibnamefont  [1]{#1}%
\providecommand \bibfnamefont [1]{#1}%
\providecommand \citenamefont [1]{#1}%
\providecommand \href@noop [0]{\@secondoftwo}%
\providecommand \href [0]{\begingroup \@sanitize@url \@href}%
\providecommand \@href[1]{\@@startlink{#1}\@@href}%
\providecommand \@@href[1]{\endgroup#1\@@endlink}%
\providecommand \@sanitize@url [0]{\catcode `\\12\catcode `\$12\catcode
  `\&12\catcode `\#12\catcode `\^12\catcode `\_12\catcode `\%12\relax}%
\providecommand \@@startlink[1]{}%
\providecommand \@@endlink[0]{}%
\providecommand \url  [0]{\begingroup\@sanitize@url \@url }%
\providecommand \@url [1]{\endgroup\@href {#1}{\urlprefix }}%
\providecommand \urlprefix  [0]{URL }%
\providecommand \Eprint [0]{\href }%
\providecommand \doibase [0]{http://dx.doi.org/}%
\providecommand \selectlanguage [0]{\@gobble}%
\providecommand \bibinfo  [0]{\@secondoftwo}%
\providecommand \bibfield  [0]{\@secondoftwo}%
\providecommand \translation [1]{[#1]}%
\providecommand \BibitemOpen [0]{}%
\providecommand \bibitemStop [0]{}%
\providecommand \bibitemNoStop [0]{.\EOS\space}%
\providecommand \EOS [0]{\spacefactor3000\relax}%
\providecommand \BibitemShut  [1]{\csname bibitem#1\endcsname}%
\let\auto@bib@innerbib\@empty
\bibitem [{\citenamefont {Uzan}(2003)}]{Uza03}%
  \BibitemOpen
  \bibfield  {author} {\bibinfo {author} {\bibfnamefont {J.-P.}\ \bibnamefont
  {Uzan}},\ }\href {\doibase 10.1103/RevModPhys.75.403} {\bibfield  {journal}
  {\bibinfo  {journal} {Rev. Mod. Phys.}\ }\textbf {\bibinfo {volume} {75}},\
  \bibinfo {pages} {403} (\bibinfo {year} {2003})}\BibitemShut {NoStop}%
\bibitem [{\citenamefont {Flambaum}(2007)}]{Fla07}%
  \BibitemOpen
  \bibfield  {author} {\bibinfo {author} {\bibfnamefont {V.~V.}\ \bibnamefont
  {Flambaum}},\ }\href {\doibase 10.1142/S0217751X07038293} {\bibfield
  {journal} {\bibinfo  {journal} {Int. J. Mod. Phys. A}\ }\textbf {\bibinfo
  {volume} {22}},\ \bibinfo {pages} {4937} (\bibinfo {year}
  {2007})}\BibitemShut {NoStop}%
\bibitem [{\citenamefont {Flambaum}(2008)}]{Fla08}%
  \BibitemOpen
  \bibfield  {author} {\bibinfo {author} {\bibfnamefont {V.~V.}\ \bibnamefont
  {Flambaum}},\ }\href {\doibase 10.1140/epjst/e2008-00817-5} {\bibfield
  {journal} {\bibinfo  {journal} {Eur. Phys. J. Sp. Top.}\ }\textbf {\bibinfo
  {volume} {163}},\ \bibinfo {pages} {159} (\bibinfo {year}
  {2008})}\BibitemShut {NoStop}%
\bibitem [{\citenamefont {Murphy}\ \emph {et~al.}(2007)\citenamefont {Murphy},
  \citenamefont {Webb},\ and\ \citenamefont {Flambaum}}]{MurWebFla07}%
  \BibitemOpen
  \bibfield  {author} {\bibinfo {author} {\bibfnamefont {M.~T.}\ \bibnamefont
  {Murphy}}, \bibinfo {author} {\bibfnamefont {J.~K.}\ \bibnamefont {Webb}}, \
  and\ \bibinfo {author} {\bibfnamefont {V.~V.}\ \bibnamefont {Flambaum}},\
  }\href {\doibase 10.1103/PhysRevLett.99.239001} {\bibfield  {journal}
  {\bibinfo  {journal} {Phys. Rev. Lett.}\ }\textbf {\bibinfo {volume} {99}},\
  \bibinfo {pages} {239001} (\bibinfo {year} {2007})}\BibitemShut {NoStop}%
\bibitem [{\citenamefont {Webb}\ \emph {et~al.}(2010)\citenamefont {Webb},
  \citenamefont {King}, \citenamefont {Murphy}, \citenamefont {Flambaum},
  \citenamefont {Carswell},\ and\ \citenamefont {Bainbridge}}]{WebKinMur10}%
  \BibitemOpen
  \bibfield  {author} {\bibinfo {author} {\bibfnamefont {J.~K.}\ \bibnamefont
  {Webb}}, \bibinfo {author} {\bibfnamefont {J.~A.}\ \bibnamefont {King}},
  \bibinfo {author} {\bibfnamefont {M.~T.}\ \bibnamefont {Murphy}}, \bibinfo
  {author} {\bibfnamefont {V.~V.}\ \bibnamefont {Flambaum}}, \bibinfo {author}
  {\bibfnamefont {R.~F.}\ \bibnamefont {Carswell}}, \ and\ \bibinfo {author}
  {\bibfnamefont {M.~B.}\ \bibnamefont {Bainbridge}},\ }\href@noop {} {}
  (\bibinfo {year} {2010}),\ \bibinfo {note} {arXiv:1008.3907v1}\BibitemShut
  {NoStop}%
\bibitem [{\citenamefont {Dzuba}\ \emph
  {et~al.}(1999{\natexlab{a}})\citenamefont {Dzuba}, \citenamefont {Flambaum},\
  and\ \citenamefont {Webb}}]{DzuFlaWeb99}%
  \BibitemOpen
  \bibfield  {author} {\bibinfo {author} {\bibfnamefont {V.~A.}\ \bibnamefont
  {Dzuba}}, \bibinfo {author} {\bibfnamefont {V.}~\bibnamefont {Flambaum}}, \
  and\ \bibinfo {author} {\bibfnamefont {J.~K.}\ \bibnamefont {Webb}},\ }\href
  {\doibase 10.1103/PhysRevA.48.546} {\bibfield  {journal} {\bibinfo  {journal}
  {Phys. Rev. Lett.}\ }\textbf {\bibinfo {volume} {82}},\ \bibinfo {pages}
  {888} (\bibinfo {year} {1999}{\natexlab{a}})}\BibitemShut {NoStop}%
\bibitem [{\citenamefont {Dzuba}\ \emph
  {et~al.}(1999{\natexlab{b}})\citenamefont {Dzuba}, \citenamefont {Flambaum},\
  and\ \citenamefont {Webb}}]{DzuFlaWeb99_2}%
  \BibitemOpen
  \bibfield  {author} {\bibinfo {author} {\bibfnamefont {V.~A.}\ \bibnamefont
  {Dzuba}}, \bibinfo {author} {\bibfnamefont {V.}~\bibnamefont {Flambaum}}, \
  and\ \bibinfo {author} {\bibfnamefont {J.~K.}\ \bibnamefont {Webb}},\ }\href
  {\doibase 10.1103/PhysRevA.48.546} {\bibfield  {journal} {\bibinfo  {journal}
  {Phys. Rev. A}\ }\textbf {\bibinfo {volume} {59}},\ \bibinfo {pages} {230}
  (\bibinfo {year} {1999}{\natexlab{b}})}\BibitemShut {NoStop}%
\bibitem [{\citenamefont {Levshakov}\ \emph
  {et~al.}(2010{\natexlab{a}})\citenamefont {Levshakov}, \citenamefont
  {Molaro}, \citenamefont {Lapinov}, \citenamefont {Reimers}, \citenamefont
  {Henkel},\ and\ \citenamefont {Sakai}}]{LevMolLap10}%
  \BibitemOpen
  \bibfield  {author} {\bibinfo {author} {\bibfnamefont {S.~A.}\ \bibnamefont
  {Levshakov}}, \bibinfo {author} {\bibfnamefont {P.}~\bibnamefont {Molaro}},
  \bibinfo {author} {\bibfnamefont {A.~V.}\ \bibnamefont {Lapinov}}, \bibinfo
  {author} {\bibfnamefont {D.}~\bibnamefont {Reimers}}, \bibinfo {author}
  {\bibfnamefont {C.}~\bibnamefont {Henkel}}, \ and\ \bibinfo {author}
  {\bibfnamefont {T.}~\bibnamefont {Sakai}},\ }\href {\doibase
  10.1103/PhysRevA.48.546} {\bibfield  {journal} {\bibinfo  {journal} {Astron.
  Astophys.}\ }\textbf {\bibinfo {volume} {512}},\ \bibinfo {pages} {A44}
  (\bibinfo {year} {2010}{\natexlab{a}})}\BibitemShut {NoStop}%
\bibitem [{\citenamefont {Levshakov}\ \emph
  {et~al.}(2010{\natexlab{b}})\citenamefont {Levshakov}, \citenamefont
  {Lapinov}, \citenamefont {Henkel}, \citenamefont {Molaro}, \citenamefont
  {Reimers}, \citenamefont {Kozlov},\ and\ \citenamefont
  {Afganova}}]{LevLapHen10}%
  \BibitemOpen
  \bibfield  {author} {\bibinfo {author} {\bibfnamefont {S.~A.}\ \bibnamefont
  {Levshakov}}, \bibinfo {author} {\bibfnamefont {A.~V.}\ \bibnamefont
  {Lapinov}}, \bibinfo {author} {\bibfnamefont {C.}~\bibnamefont {Henkel}},
  \bibinfo {author} {\bibfnamefont {P.}~\bibnamefont {Molaro}}, \bibinfo
  {author} {\bibfnamefont {D.}~\bibnamefont {Reimers}}, \bibinfo {author}
  {\bibfnamefont {M.~G.}\ \bibnamefont {Kozlov}}, \ and\ \bibinfo {author}
  {\bibfnamefont {I.~I.}\ \bibnamefont {Afganova}},\ }\href {\doibase
  10.1103/PhysRevA.48.546} {\bibfield  {journal} {\bibinfo  {journal} {Astron.
  Astophys.}\ }\textbf {\bibinfo {volume} {524}},\ \bibinfo {pages} {A32}
  (\bibinfo {year} {2010}{\natexlab{b}})}\BibitemShut {NoStop}%
\bibitem [{\citenamefont {Prestage}\ \emph {et~al.}(1995)\citenamefont
  {Prestage}, \citenamefont {Tjoelker},\ and\ \citenamefont
  {Maleki}}]{PreTjoMal95}%
  \BibitemOpen
  \bibfield  {author} {\bibinfo {author} {\bibfnamefont {J.~D.}\ \bibnamefont
  {Prestage}}, \bibinfo {author} {\bibfnamefont {R.~L.}\ \bibnamefont
  {Tjoelker}}, \ and\ \bibinfo {author} {\bibfnamefont {L.}~\bibnamefont
  {Maleki}},\ }\href {\doibase 10.1103/PhysRevLett.74.3511} {\bibfield
  {journal} {\bibinfo  {journal} {Phys. Rev. Lett.}\ }\textbf {\bibinfo
  {volume} {74}},\ \bibinfo {pages} {3511} (\bibinfo {year}
  {1995})}\BibitemShut {NoStop}%
\bibitem [{\citenamefont {Marion}\ \emph {et~al.}(2003)\citenamefont {Marion},
  \citenamefont {Pereira Dos~Santos}, \citenamefont {Abgrall}, \citenamefont
  {Zhang}, \citenamefont {Sortais}, \citenamefont {Bize}, \citenamefont
  {Maksimovic}, \citenamefont {Calonico}, \citenamefont {Gr\"unert},
  \citenamefont {Mandache}, \citenamefont {Lemonde}, \citenamefont
  {Santarelli}, \citenamefont {Laurent}, \citenamefont {Clairon},\ and\
  \citenamefont {Salomon}}]{MarPerSan03}%
  \BibitemOpen
  \bibfield  {author} {\bibinfo {author} {\bibfnamefont {H.}~\bibnamefont
  {Marion}}, \bibinfo {author} {\bibfnamefont {F.}~\bibnamefont {Pereira
  Dos~Santos}}, \bibinfo {author} {\bibfnamefont {M.}~\bibnamefont {Abgrall}},
  \bibinfo {author} {\bibfnamefont {S.}~\bibnamefont {Zhang}}, \bibinfo
  {author} {\bibfnamefont {Y.}~\bibnamefont {Sortais}}, \bibinfo {author}
  {\bibfnamefont {S.}~\bibnamefont {Bize}}, \bibinfo {author} {\bibfnamefont
  {I.}~\bibnamefont {Maksimovic}}, \bibinfo {author} {\bibfnamefont
  {D.}~\bibnamefont {Calonico}}, \bibinfo {author} {\bibfnamefont
  {J.}~\bibnamefont {Gr\"unert}}, \bibinfo {author} {\bibfnamefont
  {C.}~\bibnamefont {Mandache}}, \bibinfo {author} {\bibfnamefont
  {P.}~\bibnamefont {Lemonde}}, \bibinfo {author} {\bibfnamefont
  {G.}~\bibnamefont {Santarelli}}, \bibinfo {author} {\bibfnamefont
  {P.}~\bibnamefont {Laurent}}, \bibinfo {author} {\bibfnamefont
  {A.}~\bibnamefont {Clairon}}, \ and\ \bibinfo {author} {\bibfnamefont
  {C.}~\bibnamefont {Salomon}},\ }\href {\doibase
  10.1103/PhysRevLett.90.150801} {\bibfield  {journal} {\bibinfo  {journal}
  {Phys. Rev. Lett.}\ }\textbf {\bibinfo {volume} {90}},\ \bibinfo {pages}
  {150801} (\bibinfo {year} {2003})}\BibitemShut {NoStop}%
\bibitem [{\citenamefont {Fortier}\ \emph {et~al.}(2007)\citenamefont
  {Fortier}, \citenamefont {Ashby},\ and\ \citenamefont {{Bergquist \emph{et
  al.}}}}]{ForAshBer07}%
  \BibitemOpen
  \bibfield  {author} {\bibinfo {author} {\bibfnamefont {T.~M.}\ \bibnamefont
  {Fortier}}, \bibinfo {author} {\bibfnamefont {N.}~\bibnamefont {Ashby}}, \
  and\ \bibinfo {author} {\bibfnamefont {J.~C.}\ \bibnamefont {{Bergquist
  \emph{et al.}}}},\ }\href {\doibase 10.1103/PhysRevLett.98.070801} {\bibfield
   {journal} {\bibinfo  {journal} {Phys. Rev. Lett.}\ }\textbf {\bibinfo
  {volume} {98}},\ \bibinfo {pages} {070801} (\bibinfo {year}
  {2007})}\BibitemShut {NoStop}%
\bibitem [{\citenamefont {{Rosenband \textit{et al.}}}(2008)}]{Ros08}%
  \BibitemOpen
  \bibfield  {author} {\bibinfo {author} {\bibfnamefont {T.}~\bibnamefont
  {{Rosenband \textit{et al.}}}},\ }\href {\doibase 10.1103/PhysRevA.48.546}
  {\bibfield  {journal} {\bibinfo  {journal} {Science}\ }\textbf {\bibinfo
  {volume} {319}},\ \bibinfo {pages} {1808} (\bibinfo {year}
  {2008})}\BibitemShut {NoStop}%
\bibitem [{\citenamefont {Dzuba}\ \emph {et~al.}(2003)\citenamefont {Dzuba},
  \citenamefont {Flambaum},\ and\ \citenamefont {Marchenko}}]{DzuFlaMar03}%
  \BibitemOpen
  \bibfield  {author} {\bibinfo {author} {\bibfnamefont {V.~A.}\ \bibnamefont
  {Dzuba}}, \bibinfo {author} {\bibfnamefont {V.~V.}\ \bibnamefont {Flambaum}},
  \ and\ \bibinfo {author} {\bibfnamefont {M.~V.}\ \bibnamefont {Marchenko}},\
  }\href {\doibase 10.1103/PhysRevA.68.022506} {\bibfield  {journal} {\bibinfo
  {journal} {Phys. Rev. A}\ }\textbf {\bibinfo {volume} {68}},\ \bibinfo
  {pages} {022506} (\bibinfo {year} {2003})}\BibitemShut {NoStop}%
\bibitem [{\citenamefont {Dzuba}\ and\ \citenamefont
  {Flambaum}(2005)}]{DzuFla05}%
  \BibitemOpen
  \bibfield  {author} {\bibinfo {author} {\bibfnamefont {V.~A.}\ \bibnamefont
  {Dzuba}}\ and\ \bibinfo {author} {\bibfnamefont {V.~V.}\ \bibnamefont
  {Flambaum}},\ }\href {\doibase 10.1103/PhysRevA.72.052514} {\bibfield
  {journal} {\bibinfo  {journal} {Phys. Rev. A}\ }\textbf {\bibinfo {volume}
  {72}},\ \bibinfo {pages} {052514} (\bibinfo {year} {2005})}\BibitemShut
  {NoStop}%
\bibitem [{\citenamefont {Nguyen}\ \emph {et~al.}(2004)\citenamefont {Nguyen},
  \citenamefont {Budker}, \citenamefont {Lamoreaux},\ and\ \citenamefont
  {Torgerson}}]{NguBudLam04}%
  \BibitemOpen
  \bibfield  {author} {\bibinfo {author} {\bibfnamefont {A.~T.}\ \bibnamefont
  {Nguyen}}, \bibinfo {author} {\bibfnamefont {D.}~\bibnamefont {Budker}},
  \bibinfo {author} {\bibfnamefont {S.~K.}\ \bibnamefont {Lamoreaux}}, \ and\
  \bibinfo {author} {\bibfnamefont {J.~R.}\ \bibnamefont {Torgerson}},\ }\href
  {\doibase 10.1103/PhysRevA.69.022105} {\bibfield  {journal} {\bibinfo
  {journal} {Phys. Rev. A}\ }\textbf {\bibinfo {volume} {69}},\ \bibinfo
  {pages} {022105} (\bibinfo {year} {2004})}\BibitemShut {NoStop}%
\bibitem [{\citenamefont {Flambaum}\ and\ \citenamefont
  {Kozlov}(2007)}]{FlaKoz07}%
  \BibitemOpen
  \bibfield  {author} {\bibinfo {author} {\bibfnamefont {V.~V.}\ \bibnamefont
  {Flambaum}}\ and\ \bibinfo {author} {\bibfnamefont {M.~G.}\ \bibnamefont
  {Kozlov}},\ }\href {\doibase 10.1103/PhysRevLett.99.150801} {\bibfield
  {journal} {\bibinfo  {journal} {Phys. Rev. Lett.}\ }\textbf {\bibinfo
  {volume} {99}},\ \bibinfo {pages} {150801} (\bibinfo {year}
  {2007})}\BibitemShut {NoStop}%
\bibitem [{\citenamefont {Beloy}\ \emph {et~al.}(2010)\citenamefont {Beloy},
  \citenamefont {Borschevsky}, \citenamefont {Schwerdtfeger},\ and\
  \citenamefont {Flambaum}}]{BelBorSch10}%
  \BibitemOpen
  \bibfield  {author} {\bibinfo {author} {\bibfnamefont {K.}~\bibnamefont
  {Beloy}}, \bibinfo {author} {\bibfnamefont {A.}~\bibnamefont {Borschevsky}},
  \bibinfo {author} {\bibfnamefont {P.}~\bibnamefont {Schwerdtfeger}}, \ and\
  \bibinfo {author} {\bibfnamefont {V.~V.}\ \bibnamefont {Flambaum}},\ }\href
  {\doibase 10.1103/PhysRevA.82.022106} {\bibfield  {journal} {\bibinfo
  {journal} {Phys. Rev. A}\ }\textbf {\bibinfo {volume} {82}},\ \bibinfo
  {pages} {022106} (\bibinfo {year} {2010})}\BibitemShut {NoStop}%
\bibitem [{\citenamefont {Peik}\ and\ \citenamefont {Tamm}(2003)}]{PeiTam03}%
  \BibitemOpen
  \bibfield  {author} {\bibinfo {author} {\bibfnamefont {E.}~\bibnamefont
  {Peik}}\ and\ \bibinfo {author} {\bibfnamefont {C.}~\bibnamefont {Tamm}},\
  }\href@noop {} {\bibfield  {journal} {\bibinfo  {journal} {Europhys. Lett.}\
  }\textbf {\bibinfo {volume} {61}},\ \bibinfo {pages} {181} (\bibinfo {year}
  {2003})}\BibitemShut {NoStop}%
\bibitem [{\citenamefont {Flambaum}(2006)}]{Fla06}%
  \BibitemOpen
  \bibfield  {author} {\bibinfo {author} {\bibfnamefont {V.}~\bibnamefont
  {Flambaum}},\ }\href {\doibase 10.1103/PhysRevA.48.546} {\bibfield  {journal}
  {\bibinfo  {journal} {Phys. Rev. Lett.}\ }\textbf {\bibinfo {volume} {97}},\
  \bibinfo {pages} {092502} (\bibinfo {year} {2006})}\BibitemShut {NoStop}%
\bibitem [{\citenamefont {Litvinova}\ \emph {et~al.}(2009)\citenamefont
  {Litvinova}, \citenamefont {Feldmeier}, \citenamefont {Dobaczewski},\ and\
  \citenamefont {Flambaum}}]{LitFelDob09}%
  \BibitemOpen
  \bibfield  {author} {\bibinfo {author} {\bibfnamefont {E.}~\bibnamefont
  {Litvinova}}, \bibinfo {author} {\bibfnamefont {H.}~\bibnamefont
  {Feldmeier}}, \bibinfo {author} {\bibfnamefont {J.}~\bibnamefont
  {Dobaczewski}}, \ and\ \bibinfo {author} {\bibfnamefont {V.}~\bibnamefont
  {Flambaum}},\ }\href {\doibase 10.1103/PhysRevC.79.064303} {\bibfield
  {journal} {\bibinfo  {journal} {Phys. Rev. C}\ }\textbf {\bibinfo {volume}
  {79}},\ \bibinfo {pages} {064303} (\bibinfo {year} {2009})}\BibitemShut
  {NoStop}%
\bibitem [{\citenamefont {Chin}\ and\ \citenamefont
  {Flambaum}(2006)}]{ChiFla06}%
  \BibitemOpen
  \bibfield  {author} {\bibinfo {author} {\bibfnamefont {C.}~\bibnamefont
  {Chin}}\ and\ \bibinfo {author} {\bibfnamefont {V.~V.}\ \bibnamefont
  {Flambaum}},\ }\href {\doibase 10.1103/PhysRevLett.96.230801} {\bibfield
  {journal} {\bibinfo  {journal} {Phys. Rev. Lett.}\ }\textbf {\bibinfo
  {volume} {96}},\ \bibinfo {pages} {230801} (\bibinfo {year}
  {2006})}\BibitemShut {NoStop}%
\bibitem [{\citenamefont {Gribakin}\ and\ \citenamefont
  {Flambaum}(1993)}]{GriFla93}%
  \BibitemOpen
  \bibfield  {author} {\bibinfo {author} {\bibfnamefont {G.~F.}\ \bibnamefont
  {Gribakin}}\ and\ \bibinfo {author} {\bibfnamefont {V.~V.}\ \bibnamefont
  {Flambaum}},\ }\href {\doibase 10.1103/PhysRevA.48.546} {\bibfield  {journal}
  {\bibinfo  {journal} {Phys. Rev. A}\ }\textbf {\bibinfo {volume} {48}},\
  \bibinfo {pages} {546} (\bibinfo {year} {1993})}\BibitemShut {NoStop}%
\bibitem [{\citenamefont {Pahl}\ \emph {et~al.}()\citenamefont {Pahl},
  \citenamefont {Figgen}, \citenamefont {Borschevsky}, \citenamefont
  {Peterson},\ and\ \citenamefont {Schwerdtfeger}}]{PahFigBor11}%
  \BibitemOpen
  \bibfield  {author} {\bibinfo {author} {\bibfnamefont {E.}~\bibnamefont
  {Pahl}}, \bibinfo {author} {\bibfnamefont {D.}~\bibnamefont {Figgen}},
  \bibinfo {author} {\bibfnamefont {A.}~\bibnamefont {Borschevsky}}, \bibinfo
  {author} {\bibfnamefont {K.~A.}\ \bibnamefont {Peterson}}, \ and\ \bibinfo
  {author} {\bibfnamefont {P.}~\bibnamefont {Schwerdtfeger}},\ }\href@noop {}
  {}\bibinfo {note} {{Theo. Chem. Acc.}, in press (2011)}\BibitemShut {NoStop}%
\bibitem [{\citenamefont {Pahl}\ \emph {et~al.}(2010)\citenamefont {Pahl},
  \citenamefont {Figgen}, \citenamefont {Thierfelder}, \citenamefont
  {Peterson}, \citenamefont {Calvo},\ and\ \citenamefont
  {Schwerdtfeger}}]{PahFigThi10}%
  \BibitemOpen
  \bibfield  {author} {\bibinfo {author} {\bibfnamefont {E.}~\bibnamefont
  {Pahl}}, \bibinfo {author} {\bibfnamefont {D.}~\bibnamefont {Figgen}},
  \bibinfo {author} {\bibfnamefont {C.}~\bibnamefont {Thierfelder}}, \bibinfo
  {author} {\bibfnamefont {K.~A.}\ \bibnamefont {Peterson}}, \bibinfo {author}
  {\bibfnamefont {F.}~\bibnamefont {Calvo}}, \ and\ \bibinfo {author}
  {\bibfnamefont {P.}~\bibnamefont {Schwerdtfeger}},\ }\href {\doibase
  10.1103/PhysRevA.48.546} {\bibfield  {journal} {\bibinfo  {journal} {J. Chem.
  Phys.}\ }\textbf {\bibinfo {volume} {132}},\ \bibinfo {pages} {114301}
  (\bibinfo {year} {2010})}\BibitemShut {NoStop}%
\bibitem [{\citenamefont {Chin}\ \emph {et~al.}(2003)\citenamefont {Chin},
  \citenamefont {Kerman}, \citenamefont {Vuleti\ifmmode~\acute{c}\else
  \'{c}\fi{}},\ and\ \citenamefont {Chu}}]{ChiCheKer03}%
  \BibitemOpen
  \bibfield  {author} {\bibinfo {author} {\bibfnamefont {C.}~\bibnamefont
  {Chin}}, \bibinfo {author} {\bibfnamefont {A.~J.}\ \bibnamefont {Kerman}},
  \bibinfo {author} {\bibfnamefont {V.}~\bibnamefont
  {Vuleti\ifmmode~\acute{c}\else \'{c}\fi{}}}, \ and\ \bibinfo {author}
  {\bibfnamefont {S.}~\bibnamefont {Chu}},\ }\href {\doibase
  10.1103/PhysRevLett.90.033201} {\bibfield  {journal} {\bibinfo  {journal}
  {Phys. Rev. Lett.}\ }\textbf {\bibinfo {volume} {90}},\ \bibinfo {pages}
  {033201} (\bibinfo {year} {2003})}\BibitemShut {NoStop}%
\bibitem [{\citenamefont {Coxon}\ and\ \citenamefont
  {Hajigeorgiou}(2010)}]{CoxHaj10}%
  \BibitemOpen
  \bibfield  {author} {\bibinfo {author} {\bibfnamefont {J.~A.}\ \bibnamefont
  {Coxon}}\ and\ \bibinfo {author} {\bibfnamefont {P.~G.}\ \bibnamefont
  {Hajigeorgiou}},\ }\href@noop {} {\bibfield  {journal} {\bibinfo  {journal}
  {J. Chem. Phys.}\ }\textbf {\bibinfo {volume} {132}},\ \bibinfo {pages}
  {094105} (\bibinfo {year} {2010})}\BibitemShut {NoStop}%
\bibitem [{\citenamefont {{Saue \emph{et al.}}}()}]{DIRAC10}%
  \BibitemOpen
  \bibfield  {author} {\bibinfo {author} {\bibfnamefont {T.}~\bibnamefont
  {{Saue \emph{et al.}}}},\ }\href {http://dirac.chem.vu.nl} {\enquote
  {\bibinfo {title} {{DIRAC}, a relativistic \emph{ab initio} electronic
  structure program, release {DIRAC10}},}\ }\BibitemShut {NoStop}%
\bibitem [{\citenamefont {Hart}\ \emph {et~al.}(2007)\citenamefont {Hart},
  \citenamefont {Xu}, \citenamefont {Legere},\ and\ \citenamefont
  {Gibble}}]{HartXuLeg07}%
  \BibitemOpen
  \bibfield  {author} {\bibinfo {author} {\bibfnamefont {R.~A.}\ \bibnamefont
  {Hart}}, \bibinfo {author} {\bibfnamefont {X.}~\bibnamefont {Xu}}, \bibinfo
  {author} {\bibfnamefont {R.}~\bibnamefont {Legere}}, \ and\ \bibinfo {author}
  {\bibfnamefont {K.}~\bibnamefont {Gibble}},\ }\href {\doibase
  10.1103/PhysRevA.48.546} {\bibfield  {journal} {\bibinfo  {journal} {Nature}\
  }\textbf {\bibinfo {volume} {446}},\ \bibinfo {pages} {892} (\bibinfo {year}
  {2007})}\BibitemShut {NoStop}%
\bibitem [{\citenamefont {Herbig}\ \emph {et~al.}(2003)\citenamefont {Herbig},
  \citenamefont {Kraemer}, \citenamefont {Weber}, \citenamefont {Chin},
  \citenamefont {Nägerl},\ and\ \citenamefont {Grimm}}]{HerKraWeb03}%
  \BibitemOpen
  \bibfield  {author} {\bibinfo {author} {\bibfnamefont {J.}~\bibnamefont
  {Herbig}}, \bibinfo {author} {\bibfnamefont {T.}~\bibnamefont {Kraemer}},
  \bibinfo {author} {\bibfnamefont {T.}~\bibnamefont {Weber}}, \bibinfo
  {author} {\bibfnamefont {C.}~\bibnamefont {Chin}}, \bibinfo {author}
  {\bibfnamefont {H.-C.}\ \bibnamefont {Nägerl}}, \ and\ \bibinfo {author}
  {\bibfnamefont {R.}~\bibnamefont {Grimm}},\ }\href {\doibase
  10.1103/PhysRevA.48.546} {\bibfield  {journal} {\bibinfo  {journal}
  {Science}\ }\textbf {\bibinfo {volume} {310}},\ \bibinfo {pages} {1510}
  (\bibinfo {year} {2003})}\BibitemShut {NoStop}%
\bibitem [{\citenamefont {Ilia\v{s}}\ \emph {et~al.}(2005)\citenamefont
  {Ilia\v{s}}, \citenamefont {Jensen}, \citenamefont {Kello}, \citenamefont
  {Roos},\ and\ \citenamefont {Urban}}]{IliJenKel05}%
  \BibitemOpen
  \bibfield  {author} {\bibinfo {author} {\bibfnamefont {M.}~\bibnamefont
  {Ilia\v{s}}}, \bibinfo {author} {\bibfnamefont {H.~J.~A.}\ \bibnamefont
  {Jensen}}, \bibinfo {author} {\bibfnamefont {V.}~\bibnamefont {Kello}},
  \bibinfo {author} {\bibfnamefont {B.~O.}\ \bibnamefont {Roos}}, \ and\
  \bibinfo {author} {\bibfnamefont {M.}~\bibnamefont {Urban}},\ }\href@noop {}
  {\bibfield  {journal} {\bibinfo  {journal} {Chem. Phys. Lett.}\ }\textbf
  {\bibinfo {volume} {408}},\ \bibinfo {pages} {210} (\bibinfo {year}
  {2005})}\BibitemShut {NoStop}%
\bibitem [{\citenamefont {Faegri}(2001)}]{Fae01}%
  \BibitemOpen
  \bibfield  {author} {\bibinfo {author} {\bibfnamefont {K.}~\bibnamefont
  {Faegri}},\ }\href@noop {} {\bibfield  {journal} {\bibinfo  {journal} {Theor.
  Chim. Acta}\ }\textbf {\bibinfo {volume} {105}},\ \bibinfo {pages} {252}
  (\bibinfo {year} {2001})}\BibitemShut {NoStop}%
\bibitem [{\citenamefont {Casimir}\ and\ \citenamefont
  {Polder}(1948)}]{CasPol48}%
  \BibitemOpen
  \bibfield  {author} {\bibinfo {author} {\bibfnamefont {H.~B.~G.}\
  \bibnamefont {Casimir}}\ and\ \bibinfo {author} {\bibfnamefont
  {D.}~\bibnamefont {Polder}},\ }\href {\doibase 10.1103/PhysRev.73.360}
  {\bibfield  {journal} {\bibinfo  {journal} {Phys. Rev.}\ }\textbf {\bibinfo
  {volume} {73}},\ \bibinfo {pages} {360} (\bibinfo {year} {1948})}\BibitemShut
  {NoStop}%
\bibitem [{\citenamefont {London}(1930)}]{Lon30}%
  \BibitemOpen
  \bibfield  {author} {\bibinfo {author} {\bibfnamefont {F.}~\bibnamefont
  {London}},\ }\href {\doibase 10.1103/PhysRevA.48.546} {\bibfield  {journal}
  {\bibinfo  {journal} {Z. Phys.}\ }\textbf {\bibinfo {volume} {63}},\ \bibinfo
  {pages} {245} (\bibinfo {year} {1930})}\BibitemShut {NoStop}%
\bibitem [{\citenamefont {Power}(2001)}]{Pow01}%
  \BibitemOpen
  \bibfield  {author} {\bibinfo {author} {\bibfnamefont {E.}~\bibnamefont
  {Power}},\ }\href {\doibase 10.1103/PhysRevA.48.546} {\bibfield  {journal}
  {\bibinfo  {journal} {Eur. J. Phys.}\ }\textbf {\bibinfo {volume} {22}},\
  \bibinfo {pages} {453} (\bibinfo {year} {2001})}\BibitemShut {NoStop}%
\bibitem [{\citenamefont {Derevianko}\ \emph {et~al.}(1999)\citenamefont
  {Derevianko}, \citenamefont {Johnson}, \citenamefont {Safronova},\ and\
  \citenamefont {Babb}}]{DerJohSaf99}%
  \BibitemOpen
  \bibfield  {author} {\bibinfo {author} {\bibfnamefont {A.}~\bibnamefont
  {Derevianko}}, \bibinfo {author} {\bibfnamefont {W.~R.}\ \bibnamefont
  {Johnson}}, \bibinfo {author} {\bibfnamefont {M.~S.}\ \bibnamefont
  {Safronova}}, \ and\ \bibinfo {author} {\bibfnamefont {J.~F.}\ \bibnamefont
  {Babb}},\ }\href {\doibase 10.1103/PhysRevLett.82.3589} {\bibfield  {journal}
  {\bibinfo  {journal} {Phys. Rev. Lett.}\ }\textbf {\bibinfo {volume} {82}},\
  \bibinfo {pages} {3589} (\bibinfo {year} {1999})}\BibitemShut {NoStop}%
\bibitem [{\citenamefont {Pershina}\ \emph {et~al.}(2008)\citenamefont
  {Pershina}, \citenamefont {Borschevsky}, \citenamefont {Eliav},\ and\
  \citenamefont {Kaldor}}]{PerBorEli08}%
  \BibitemOpen
  \bibfield  {author} {\bibinfo {author} {\bibfnamefont {V.}~\bibnamefont
  {Pershina}}, \bibinfo {author} {\bibfnamefont {A.}~\bibnamefont
  {Borschevsky}}, \bibinfo {author} {\bibfnamefont {E.}~\bibnamefont {Eliav}},
  \ and\ \bibinfo {author} {\bibfnamefont {U.}~\bibnamefont {Kaldor}},\ }\href
  {\doibase 10.1021/jp8061306} {\bibfield  {journal} {\bibinfo  {journal} {J.
  Phys. Chem. A}\ }\textbf {\bibinfo {volume} {112}},\ \bibinfo {pages} {13712}
  (\bibinfo {year} {2008})}\BibitemShut {NoStop}%
\bibitem [{\citenamefont {Lim}\ \emph {et~al.}(2005)\citenamefont {Lim},
  \citenamefont {Schwerdtfeger}, \citenamefont {Metz},\ and\ \citenamefont
  {Stoll}}]{LimSchMet05}%
  \BibitemOpen
  \bibfield  {author} {\bibinfo {author} {\bibfnamefont {S.~I.}\ \bibnamefont
  {Lim}}, \bibinfo {author} {\bibfnamefont {P.}~\bibnamefont {Schwerdtfeger}},
  \bibinfo {author} {\bibfnamefont {B.}~\bibnamefont {Metz}}, \ and\ \bibinfo
  {author} {\bibfnamefont {H.}~\bibnamefont {Stoll}},\ }\href {\doibase
  10.1103/PhysRevLett.82.3589} {\bibfield  {journal} {\bibinfo  {journal} {J.
  Chem. Phys.}\ }\textbf {\bibinfo {volume} {122}},\ \bibinfo {pages} {104103}
  (\bibinfo {year} {2005})}\BibitemShut {NoStop}%
\bibitem [{\citenamefont {Amini}\ and\ \citenamefont
  {Gould}(2003)}]{AmiJasGou03}%
  \BibitemOpen
  \bibfield  {author} {\bibinfo {author} {\bibfnamefont {J.~M.}\ \bibnamefont
  {Amini}}\ and\ \bibinfo {author} {\bibfnamefont {H.}~\bibnamefont {Gould}},\
  }\href {\doibase 10.1103/PhysRevLett.91.153001} {\bibfield  {journal}
  {\bibinfo  {journal} {Phys. Rev. Lett.}\ }\textbf {\bibinfo {volume} {91}},\
  \bibinfo {pages} {153001} (\bibinfo {year} {2003})}\BibitemShut {NoStop}%
\end{thebibliography}

%

\end{document}